\documentclass[preprint,showpacs,preprintnumbers,amsmath]{revtex4-1}

\usepackage{graphicx}

\begin{document}


\title{Rational Design of Half-Metallic Heterostructures}

\author{William H. Butler}
\affiliation{Center for Materials for Information Technology and Department of Physics $\&$ Astronomy\\ University of Alabama, Box 870209, Tuscaloosa, Alabama 35487, USA}
\author{Claudia K.A. Mewes}
\affiliation{Center for Materials for Information Technology and Department of Physics $\&$ Astronomy\\ University of Alabama, Box 870209, Tuscaloosa, Alabama 35487, USA}
\author{Chunsheng Liu}
\affiliation{Center for Materials for Information Technology and Department of Physics $\&$ Astronomy\\ University of Alabama, Box 870209, Tuscaloosa, Alabama 35487, USA}
\author{Tianyi Xu}
\affiliation{Center for Materials for Information Technology and Department of Physics $\&$ Astronomy\\ University of Alabama, Box 870209, Tuscaloosa, Alabama 35487, USA}

\date{\today}

\begin{abstract}
We  present   a  rational  approach  to  the   design  of  half-metallic
heterostructures  which  allows the  design  of  an  infinite number  of
half-metallic heterostructures. The wide  range of materials that can be
made  half-metallic using  our approach  makes it  possible  to engineer
materials  with  tunable  characteristic  properties,  for  example  low
intrinsic  magnetic  damping,  small  magnetic moment  or  perpendicular
anisotropy. We  demonstrate the proposed  design scheme for a  series of
transition metal heterostructures based on the B2 crystal structure.
\end{abstract}

\pacs{75.70.-i,  75.47.De,  75.47.Np,  72.25.Ba,  75.70.Cn}

\maketitle 

A ferromagnet  can be viewed  as two different  materials simultaneously
occupying the same space; one material with the majority-spin electronic
structure,  the   other  material  with   the  minority-spin  electronic
structure.  The most extreme  situation is represented by half-metals in
which the electronic structure of one  of the spin channels is that of a
metal while that of the other is an insulator or semiconductor.  Because
only  one  of  the  spin  channels  of a  half-metal  is  conductive,  a
half-metal  can in  principle generate  a fully  spin-polarized current.
Therefore half-metals  are ideal for many present  and future spintronic
applications injecting  spin-dependent currents into  semiconductors and
such on further improving  giant magnetoresistive magnetic field sensors
for hard  drives and other  applications. They may also  be advantageous
for  a new  kind of  solid state  magnetic  memory, spin-transfer-torque
magnetic  random access  memory  (STT-MRAM) that  has  the potential  to
become a new and universal memory combining dense data storage with fast
read  and write  while being  non-votile.   Although a  large number  of
half-metals   have  been   predicted   through  first-principles   based
electronic                     structure                    calculations
\cite{deGroot1983,Schwarz1986,Galanakis2006,Wurmehl2006,Kandpal2007}
these  ``discoveries" appear  to occur  mainly through  serendipity.  In
this  paper   we  describe  a   rational  approach  to  the   design  of
half-metallic heterostructures.   This approach allows the  design of an
infinite number of half-metallic heterostructures and shows how they can
be  optimized for  important applications  of intense  current practical
interest.  Optimization might involve controlling the magnetization, the
mangnetic  damping  or  the  magnetic  anisotropy  while  maintaing  the
half-metallic feature.  It might  also involve optimizing the electronic
structure in the  vicinity of the gap to  increase the spin polarization
at finite temperature.

\emph{Creating a Gap.} - In order to make a half-metal one needs to create a band gap in either the majority or minority density of states and then one must contrive to place the Fermi energy into this gap.  One way to place a gap in the middle of a band or band complex is to apply a theorem which for brevity we call the ``gap theorem".  Consider a system with two sublattices, which for definiteness will be called $A$ and $B$.  The electronic structure of transition metal alloys can be well represented by a tight-binding Hamiltonian.  The gap theorem states that if the tight-binding representation of the system has a set of atomic orbitals with on-site energies $E_A$ on the $A$ sites and $E_B$ on the $B$ sites and if the $A$ orbitals only interact with the $B$ oribitals, then there will be no states between $E_A$ and $E_B$.  There may be several orbitals on each site, but all of the orbitals on the same site must have the same on-site energy for the theorem to apply. The origin of the theorem can be easily understood by considering a chain of atoms with a single orbital per atom and nearest neighbor interactions (defined by hopping parameter, $w$). It is assumed that alternate atoms have on-site energies $E_A$ and $E_B$ (see Figure 1).  The system Hamiltonian can then be written as:
\begin{eqnarray}
\hat{H}=\sum_{n\ odd}E_A\hat{c}_n^\dag\hat{c}_n+\sum_{n\ even}E_B\hat{c}_n^\dag\hat{c}_n+\\
\sum_{n}w(\hat{c}_n^\dag\hat{c}_{n+1}+\hat{c}_n\hat{c}_{n+1}^\dag). \nonumber
\label{eq:one}
\end{eqnarray}
The density states $n(E)$ can be calculated easily using elementary means and is given by
\begin{eqnarray}
n(E)=\frac{1}{w\pi}\mbox{Im}\left[\frac{x_A+x_B}{\sqrt{x_A^2x_B^2-4x_Ax_B}}\right],
\label{eq:two}
\end{eqnarray}
where $x_A=(E-E_A)/w$ and $x_B=(E-E_B)/w$. The density of electronic states vanishes if $x_Ax_B<0$ because the quantity inside the square brackets will be real. Thus the density of states vanishes if $(E-E_A)(E-E_B)<0$, or equivalently if $E_A<E<E_B$.  The system will have a gap extending from $E_A$ to $E_B$.

The gap is independent of the inter-site interaction, $w$, which allows us to easily extend the result to three dimensional systems such as the B2 (or CsCl) system which can be viewed as atomic layers (in this case along (001)) in which, because of the assumption of only nearest neighbor interactions, there are no intra-layer interactions.  This means that a lattice Fourier transform in the plane of the layers (perpendicular to (001)) yields expressions formally identical to the expression above, but with $w$ and hence $n(E)$ dependent on $k_{\parallel}$, the wave vector parallel to the (001) planes.  Since the gap between $E_A$ and $E_B$ is not dependent on $k_{\parallel}$, the gap will remain even after integration over $k_{\parallel}$.

The theorem can also be applied to a set of degenerate orbitals with different symmetries.  As long as the interactions are restricted to nearest neighbors there will be a gap between the onsite energy for the A sites and the onsite energy for the B sites.  This is illustrated in Figure 2 for a model B2 system with only d-electrons.  This calculation is for a non-orthogonal tight-binding model for a B2 alloy with 5 d-states (of correct symmetry) on each lattice site.  Note that the gap occurs for half-filling, 5 states per two atom cell fall below the gap and 5 above.  Extension of the arguments of the preceding two paragraphs from the orthogonal tight-binding model to the non-orthogonal case is straight-forward.

\emph{Placing the gap at the Fermi energy} - In the following we shall use the gap theorem as an aid in designing half-metallic alloys and compounds. To apply the gap theorem to ``real"  transition metal alloys and compounds one must recognize that the correspondence between real transition metal alloys and the model for which the gap theorem applies is only approximate. One approximation involved in applying the gap theorem to real materials is the restriction to nearest neighbor interactions. The interactions in real materials often extend over greater distances than nearest neighbors and the s- and d- orbitals are not precisely degenerate.  The p-orbitals are usually sufficiently far above the Fermi energy that their influence on the lower lying states can be viewed as a higher order effect.

As an example of the application of the gap theorem, consider the density of states (DOS) of bcc transition metals.  The DOS of such systems is well known to have a pronounced minimum when the d-band is approximately half filled. Less well known, but easily demonstrated, is that the minimum becomes much deeper and wider when corner and body atoms are different.  This can be understood on the basis of the gap theorem. Consider panels (a-c) of Figure 3 which show a series of DOS for B2 transition metal alloys calculated using first principles with the generalized gradient approximation using the Vienna ab-initio simulation package \cite{Kresse1997}. If these systems had only d-states and if interactions only extended to nearest neighbors, there would be a gap between the onsite energies of the $A$ atoms and the onsite energies of the $B$ atoms. Rather than a gap, there is a deep, wide minimum in the minority density of states at half-filling which we refer to as a pseudo-gap. For each system shown in Figure 3 (a-c) and for many other B2 transition metal systems there is a deep pseudo-gap at 3 electrons per atom.

For each of the systems shown in Fig. 3 (a-c) the large pseudo-gap allows the system to lower its energy by shifting electrons from one spin channel to the other, to place the Fermi energy $(E_F)$ for one spin channel in the pseudo-gap.  Thus the second major task in making a half-metal (placing the gap at the Fermi energy) is often accomplished by nature as it contrives to lower the system energy. It should be realized that a number of other factors come into play when the system places the Fermi energy within the pseudo-gap. Shifting electrons between minority to majority changes the magnetic moment and  the band energy. This band energy increase can be partially compensated by Hund's rule exchange energy which is lowered by maximizing the magnetic moment. Furthermore  any energy cost from shifting the bands to increase the moment can be mitigated by increasing the volume. This narrows the bands and reduces the energy cost of increasing the magnetic moment.

The system can further lower its energy if it enhances the pseudo-gap by increasing the difference in the number of electrons for A and B atoms for the spin channel in which the Fermi energy falls in the gap. According to our first-principles calculations, this happens for many B2 3d transition metal alloys. This enhancement effect leads to an approximate rule of thumb for these alloys: in the pseudo-gapped spin channel, the approximate electron count on alternate layers is 4 and 2 \cite{Adams}. The 4-2 rule results from a compromise: the system would like to lower its energy by making the electron count as different as possible, but is constrained by the requirements that the average must be 3, that there are only 5 d-states in total, and that the early 3d transition metals are difficult to polarize.

The examples of Fe-Ti and Co-Mn are instructive. For Fe-Ti, the``4-2 rule" forces Fe to divide its 8 electrons as 4 up and 4 down, while Ti also divides its electrons equally as 2 up and 2 down. For this system, the 4-2 rule is sufficiently robust to cancel the magnetic moment of Fe and make a transition metal alloy that is $50\%$ Fe into a nonmagnetic semimetal. For Co-Mn, Co easily polarizes as 5 up, 4 down while Mn produces a large 3 $\mu_B$ moment to provide 5 up and 2 down. Thus the majority channel acts (approximately) like a pure bcc material (similar to pure Fe) while the minority channel has a pseudo-gap.

One can maintain the pseudo-gap while combining different B2 systems with relative impunity by layering along the 001 direction - as long as the atoms can realistically observe the 4-2 rule, they will attempt to maintain the pseudo-gap. More precisely, the energy reduction achieved by placing $E_F$ of one of the spin channels in the pseudo-gap will provide the driving force to produce the 4-2 rule. This is illustrated by Fig. 3 (c), which shows the DOS for an (001) layered bcc alloy consisting of Co-Mn-Fe-Ti atomic layers. The minimum of the pseudo-gap in the minority DOS is at precisely 12 electrons/4 atom cell.


\emph{Converting the pseudo-gap into a real gap} - The pseudo-gap is not a true gap primarily because the next nearest neighbor (NNN) interactions in these materials do not vanish. The NNN interactions are larger for the transition metal atom with higher on-site energy (fewer valence electrons) because they have more extended d-orbitals, and for the $e_g$ symmetry states because they have the strongest overlaps in the NNN directions. This can be seen in Fig. 3 (a) which shows the Mn $e_g$ DOS. The B2 pseudo-gaps can be converted into real gaps by substituting some of the higher on-site energy atoms (e.g. Mn and Ti here) with atoms that do not have d-states. The decreased opportunity for next-nearest neighbor hybridization will narrow the bands and prevent the DOS from tailing into the gap. It is important, however, that the substituted atom is able to sustain the 4-2 rule - thus Si and Ge are good substitutions for the atom with 2 minority electrons. Figure 3 (d)-(e) shows that the three pseudo-gapped systems (Fig. 3 (a)-(c) ) can be converted into half-metals by adding Si.

Considering these ideas further, an infinite number of layered half-metallic structures can be obtained. Note that the Co$_2$MnSi and Fe$_2$TiSi are consistent with the L2$_1$ or ``full" Heusler crystal structure \cite{Galanakis2002,Felser2003,Galanakis2006,Felser2007}. For the L2$_1$ structures Co$_2$MnSi and Co$_2$MnGe it has been observed that for a large number of insertions of single  or double atomic layers of various transition metals or combinations of transition metals and nontransition metals the band gap in the minority spin channel remains \cite{Culbert2008}. The well-known L2$_1$ Heuslers are a subset of the proposed class of half-metals. We emphasize however that the L2$_1$ structure is not required for these materials to be half-metallic. Calculations on supercells of B2 CoMn demonstrate this.  In these calculations, random substitution of Mn by Si produces half-metallicity for Si concentrations from $<12\%$ to $>33\%$ as shown in Fig. 4.

The extremely wide range of materials that we believe can be made half-metallic opens interesting new possibilities. Some of these materials can be made half-metallic or nearly so with very small or zero moment. The zero moment semimetals (e.g. systems based on FeTi or FeV), may make excellent spacer materials. The very low moment half-metals may also be useful for spin-torque switching.  By making heterostructues, we can generate uniaxial anisotropy and perpendiulcar magnetocrystalline anisotropy. The tremendous flexibility in materials also makes it possible to engineer materials with different characteristic properties, for example low intrinsic magnetic damping \cite{Liu2009,Lee2009} which is very important for reducing the critical switching current for spin-torque switching and for increasing the power output of spin-torque oscillators.

\emph{Other Families of Half-Metals} - In addition to the B2 structrure, other systems composed of two atomic species in which the A atoms have only B atoms as nearest neighbors (and vice versa) include systems such the rock salt or B1 structure and the Zincblende or B3 structure.   These structures are typically favored by materials with ionic bonds for which the gap theorem usually yields gaps in both channels, i.e. insulators.  The transition metal arsenides, VAs, CrAs, and MnAs, however, are predicted to be half-metals in the B3 phase \cite{Shirai2001,Shirai2003}.  It is also possible to generalize the gap theorem so that it can be used for more than two types of atoms and to place gaps at other points in the band than the mid-point.

In summary we have presented a rational approach to the design of half-metallic heterostructures. This scheme allows the design of an infinite number of half-metallic heterostructures. The underlying physical mechanism has been identified as the so called gap theorem in combination with a 4-2 rule. The design scheme has been demonstrated for a series of B2 transition metal alloys but can be applied to other systems.  The rational design scheme allows a wide range of materials to be made half-metallic and allows the optimization of those half-metallic structures for particular applications.

\begin{acknowledgments}
This work was partly supported by NSF-DMR MRSEC 0213985.
\end{acknowledgments}


\clearpage
\begin{figure}
\includegraphics[width=0.7\textwidth]{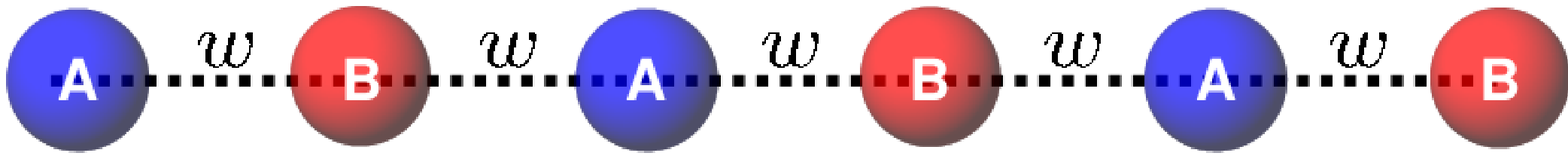}
\caption{One dimensional alternating atomic chain with only nearest neighbor interactions.}
\end{figure}

\clearpage
\begin{figure}
\includegraphics[width=0.7\textwidth]{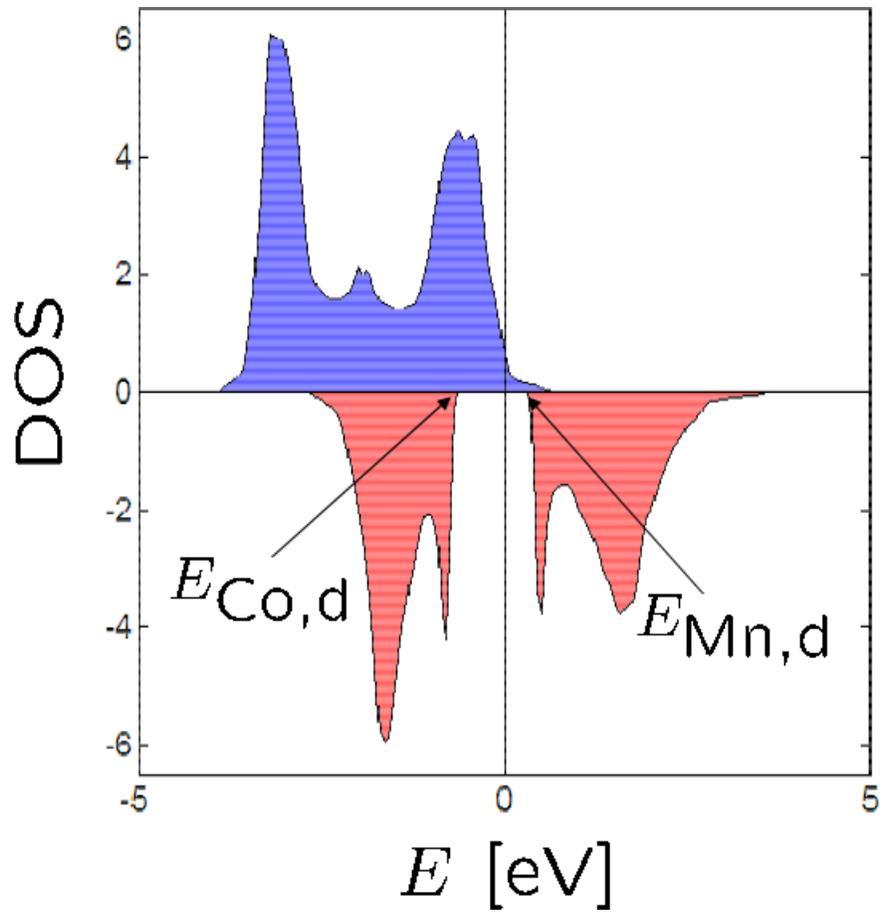}
\caption{Density of states for a B2 lattice (CoMn) including only d-states. The interaction range is restricted to nearest neighbors.}
\end{figure}

\clearpage
\begin{figure}
\includegraphics[width=0.7\textwidth]{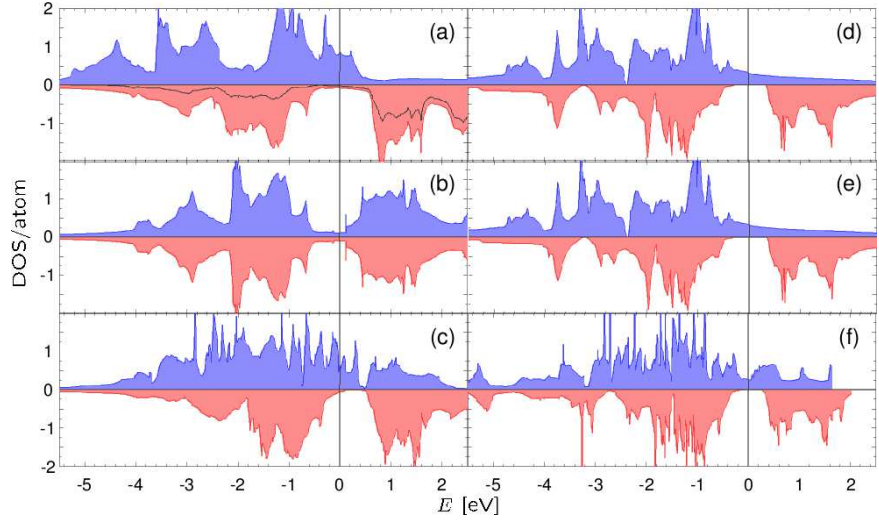}
\caption{Density of states for (a) CoMn, (b) FeTi, (c) CoMnFeTi, (d) Co$_2$MnSi, (e) Fe$_2$TiSi and (f) Co$_2$MnSiFe$_2$TiSi, positive energies: majority spins, negative energies: minority spins, additional line in subfigure (a) shows the Mn $e_g$ DOS.}
\end{figure}

\clearpage
\begin{figure}
\includegraphics[width=0.7\textwidth]{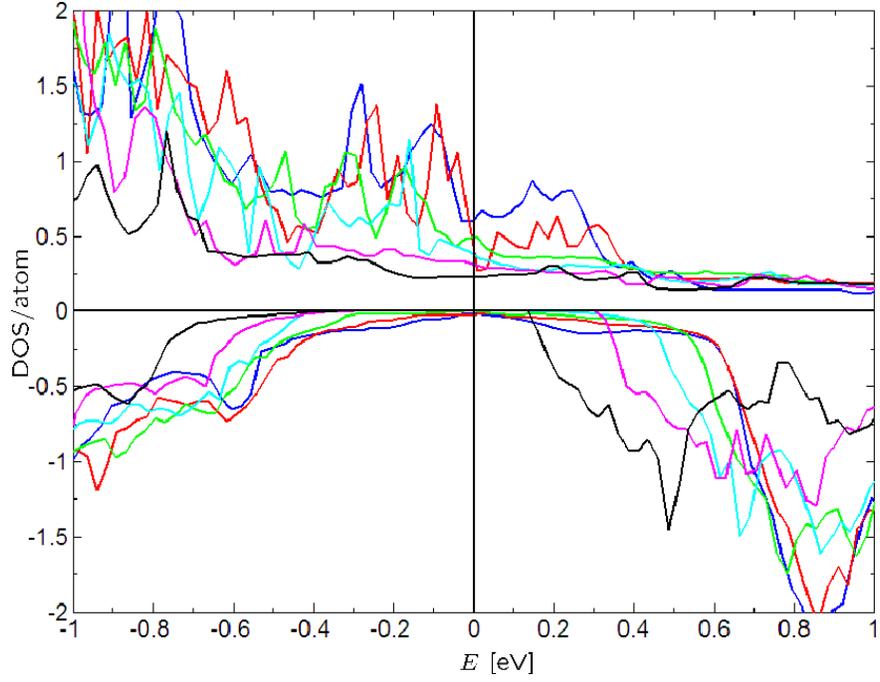}
\caption{Density of states for a 16 atom super cell of B2 Co$_8$Mn$_{8-n}$Si$_n$ with $n$ atoms of Si replacing Mn, blue, red, green, cyan, magenta and black correspond to $n=0,1,\ldots,5$ respectively.}
\end{figure}

\end{document}